# Thermal expansion and the glass transition

**Peter Lunkenheimer[1✉], Alois Loidl[1], Birte Riechers[2,3], Alessio Zaccone[4,5,6] and Konrad Samwer[7]**

[1] Experimental Physics V, Center for Electronic Correlations and Magnetism, University of Augsburg, 86135 Augsburg, Germany
[2] Bundesanstalt für Materialforschung und -prüfung, Unter den Eichen 87, 12205 Berlin, Germany
[3] Glass and Time, Department of Science and Environment, Roskilde University, DK-4000 Roskilde, Denmark
[4] Department of Physics "A. Pontremoli", University of Milan, Via Celoria 16, 20133 Milan, Italy
[5] Department of Chemical Engineering and Biotechnology, University of Cambridge, Cambridge CB3 0AS, U.K.
[6] Cavendish Laboratory, University of Cambridge, Cambridge, CB3 0HE, U.K.
[7] 1. Physikalisches Institut, University of Goettingen, Germany

**Melting is well understood in terms of the Lindemann criterion, essentially stating that crystalline materials melt when the thermal vibrations of their atoms become such vigorous that they shake themselves free of the binding forces. However, how about another common type of solids: glasses, where the nature of the solid-liquid crossover is highly controversial? The Lindemann criterion implies that the thermal expansion coefficients $\alpha$ of crystals are inversely proportional to their melting temperatures. Here we find that, unexpectedly, $\alpha$ of glasses decreases much stronger with increasing glass-transition temperature $T_g$ marking the liquid-solid crossover in this material class. However, scaling $\alpha$ by the fragility $m$, a measure of particle cooperativity, restores the proportionality, i.e., $\alpha/m \propto 1/T_g$. Obviously, for a glass to become liquid, it is not sufficient to simply overcome the interparticle binding energies. Instead, more energy has to be invested to break up the typical cooperative particle network which is considered a hallmark feature of glassy materials. Surprisingly, $\alpha$ of the liquid phase reveals similar anomalous behaviour and is universally enhanced by a constant factor of ~3. The found universalities allow estimating glass-transition temperatures from thermal expansion and vice versa.**

Many materials in technology and nature are glasses, disordered materials that are solid but lack the periodicity of the crystalline lattice[1,2]. This not only includes the common silica-based transparent materials used for windows, glass fibres, etc., but also many polymers and bio-derived materials, various solid-state electrolytes, supercooled molecular liquids and even amorphous metals. This state of matter is usually prepared by cooling a liquid sufficiently fast to avoid crystallization[1,2,3]. Below the melting temperature $T_m$, then a so-called supercooled liquid is formed first, before the material becomes a glass below the glass-transition temperature $T_g < T_m$. The latter marks the boundary between liquid and solid which usually is defined at a viscosity value of $10^{12}$ Pas. However, in contrast to crystallization, the solidification at $T_g$ occurs smoothly, i.e. without a discontinuous jump of the viscosity. Below $T_g$, most physical quantities of a glass former exhibit a crossover to weaker temperature dependence, i.e. a jump in their derivatives, at first glance reminding of a second-order phase transition. This is also the case for the volume (cf. Fig. 1a), respectively the thermal expansion, treated in the present work.

Although mankind is using supercooling for the preparation of glasses since millennia, there is no consensus on the true nature of the glass transition[1,2,3,4,5]. The temperature of the mentioned crossover depends on the cooling rate, clearly excluding a canonical phase transition. Instead, it is commonly assumed that the liquid falls out of thermodynamic equilibrium at the glass transition, which happens just at $T_g$ for a typical cooling rate of 10 K/min. Nevertheless, various competing theoretical approaches assume that an underlying, "hidden" phase transition at a temperature above or below $T_g$ may in fact govern the crossover between liquid and glass[2,4,6,7]. Alternatively, it simply could be a purely kinetic phenomenon[4,8,9].

In contrast, the transition between the liquid and solid states via crystallization and melting is much better understood[10], in particular in terms of the basic ideas behind the Lindemann criterion[11,12]. The latter predicts that melting occurs when the root mean-square (rms) displacement of particles due to thermal vibrations exceeds a certain percentage of the interparticle spacing[12], often reported to be roughly of the order of 10 % (refs. [13,14,15]). It is nowadays well established that these vibrations take place within potential wells whose asymmetry gives rise to thermal expansion. The higher the melting temperature, the deeper the well, and, hence, the steeper the slope of its attractive part should be (cf. Fig. 1b). As this slope $s$ is related to the thermal-expansion coefficient $\alpha_c$ of a crystalline material, one can expect less expansion for materials with higher $T_m$. Making the reasonable approximations that $T_m \propto U_0$ (with $U_0$ the depth of the well) and that $1/\alpha_c \propto s \propto U_0$ (ref. [16]; see Supplementary Note 1 for a more detailed discussion), one arrives at:

$$\alpha_c T_m = const. \qquad (1)$$

Here $\alpha_c$ is defined as the relative volume change at constant pressure $p$, namely $\alpha_c = 1/V\,(\partial V/\partial T)_p$. Indeed, such a relation was suggested to be directly related to the Lindemann criterion[14,17] (see also Supplementary Note 1).

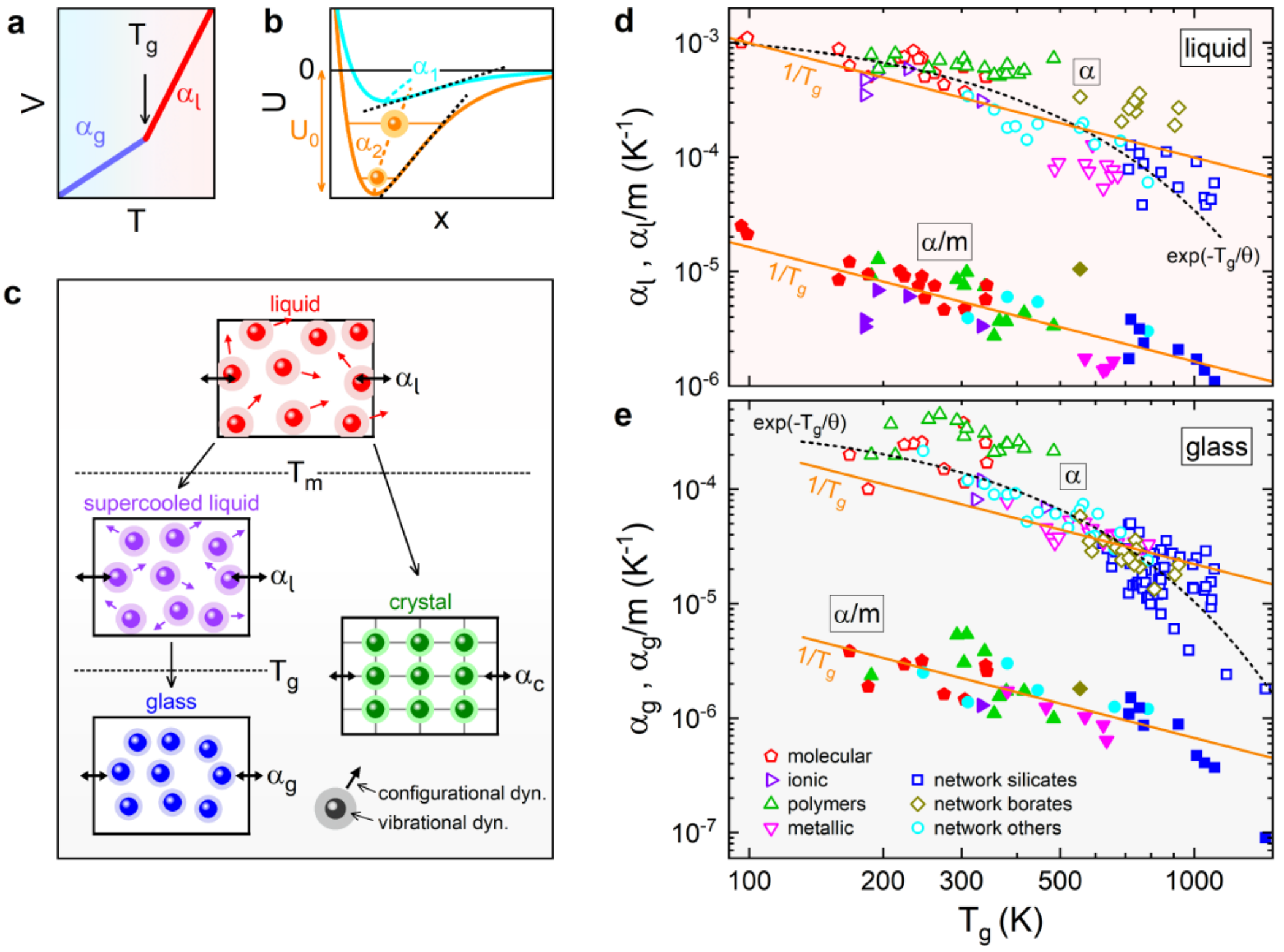

**Fig. 1 Contributions to the thermal expansion and its correlations with the glass-transition temperature. a**, Schematic plot of the temperature variation of the volume around the glass transition. **b**, Schematic plot of the asymmetric pair potential giving rise to thermal expansion in a solid. Two potentials for two different binding energies (depths of minima) are shown. The dotted black lines show the slopes at the attractive parts of the potentials which are smaller for lower binding energy. The dashed lines indicate the average location of the particle which shifts to the right (larger interparticle distance) for higher temperature, leading to thermal expansion. For the deeper potential, the particle position is shown for two temperatures. **c**, Schematic presentation of the different contributions to the thermal expansion of liquids, glasses and crystals: the vibrational dynamics is indicated by the shaded areas around the spheres, representing the atoms or molecules of the material. The additional configurational dynamics in the liquid phases is indicated by single-headed arrows. The double-headed arrows illustrate the resulting thermal expansion. **d**,**e**, Double-logarithmic plot of the experimentally-determined thermal volume-expansion coefficients $\alpha_g$ in the glass phase (**d**) and of $\alpha_l$ in the liquid phase (**e**) versus the glass-transition temperature $T_g$ for a large variety of glass formers belonging to different material classes (see Supplementary Table 1 for detailed information on all materials and values and the corresponding references). In addition to the bare expansion coefficients (open symbols), the figure also provides the $\alpha$ values divided by the fragility parameter $m$ (filled symbols) being a measure of cooperative dynamics. The solid lines show linear fits with slope -1, based on all data points for each phase, except for $\alpha_l$ of the borates. The dashed lines represent fits with $\alpha \propto \exp(-T_g/\theta)$ with the same $\theta \approx 270$ K for both data sets. Note that the ordinates of (**d**) and (**e**) were adjusted to achieve the same decades/cm ratio.

In crystalline solids, the ordered structure melts at the melting temperature and in glasses the rigid disordered structure dissolves above the glass-transition temperature. Thus, it seems natural that these two phenomena have a common basis, specifically having in mind that for many glasses the relation $T_g \approx 2/3\, T_m$ holds[18,19,20,21,22] (but also exceptions were reported[23]). In light of a possible Lindemann-like criterion for the glass-liquid transition considered, e.g., in refs. [4,21,22,24,25,26,27,28], in analogy to crystals one thus may expect the relation

$$\alpha_g T_g = const. \quad (2)$$

In general, the thermal expansion is of fundamental importance, defining universal quantities such as the Grüneisen parameter or the Prigogine-Defay ratio[1,12,29]. It also reflects the occurrence of different dynamic processes in glasses.[30] The change of slope of $V(T)$ at $T_g$ (Fig. 1a) is one of the most paradigmatic characteristics of the glass transition[18,31,32]. The thermal-expansion coefficient in liquids, $\alpha_l$, is by about a factor of 1.5 - 4 higher than in solids[32,33,34,35]. It is well established that $\alpha_l$ contains two contributions: a vibrational one, also present in the solid state, and an additional configurational one, being caused by the translational motions of the particles that also give rise to the viscous flow defining a liquid[22,33,34,36] (see schematic representation in Fig. 1c). The vibrational contribution arises from the anharmonic interparticle potential and dominates the thermal expansions of crystals and glasses, which mostly are of similar magnitude.

Interestingly, Stillinger and co-workers suggested a Lindemann-like *freezing* criterion for liquids[37,38,39]: Based on molecular-dynamics simulations, they found that melts freeze if the rms particle displacement falls below about one-half of the interparticle spacing. In analogy to equations (1) and (2), related to the Lindemann *melting* criterion, one thus could naively expect

$$\alpha_l T_g = const. \tag{3}$$

with $\alpha_l$ the expansion coefficient of the liquid. However, $\alpha_l$ is believed to be governed by additional configurational motions instead of the vibrations exclusively considered in the Lindemann scenario. Therefore, deviations from such a correlation, if present at all, may be expected. Nevertheless, in ref. [40] such a relation was predicted, based on theoretical considerations. Moreover, within the framework of the recently developed Krausser-Samwer-Zaccone (KSZ) model[41], equation (3) should also be approximately valid. In this model, a steepness parameter $\lambda$ determines the repulsive part of the inter-particle potential and is a proxy for the chemistry-dependent bonding. If $\alpha_l T_g$ is independent of $\lambda$, *i.e* of chemistry, KSZ predict an approximately linear relation between the fragility index $m$ (refs. [42,43]) and $\lambda$. Indeed, this prediction was recently found to be fulfilled for a large variety of glass formers[44], in accord with equation (3).

In literature there are some reports on, partly contradicting, correlations of $T_g$ with the thermal expansion or with $\Delta\alpha$, the jump of $\alpha$ at $T_g$, namely: $\Delta\alpha T_g = const.$[45,46], $\Delta\alpha T_g \propto T_g$ (ref. [47]), $\alpha_g T_g^2 = const.$[20], and $\alpha_l T_g = const.$[40,45] (equation (3)). However, they all were found for specific classes of glass-formers only and the overall data base was limited. In contrast, in the present work, using data on more than 200 materials from literature (see Supplementary Table 1), we check for such correlations across very different classes of glass formers.

If equations (2), (3) or alternative universal relations hold, $\alpha$ measured in a glass or liquid would allow to predict glass-transition temperatures, without any knowledge of microscopic pair-potential parameters. At the same time, one could gain insight into the universality of configurational contributions to the thermal expansion at $T > T_g$ and concerning the relevance of a Lindemann-like mechanism for the glass transition. In any case, the explanation of a possible universal relationship of $\alpha$ and $T_g$ would represent a severe benchmark for any model of the glass transition.

## Experimental data and analysis

The values of $\alpha_g$, $\alpha_l$ and $T_g$ used in the present work are listed in Supplementary Table 1 and details on their selection and reliability are provided in the Supplementary Notes 2 and 3. The included materials can be classified as molecular glass formers (alcohols, van-der-Waals bonded and other systems), polymers, ionic glass formers (including ionic liquids and melts), metallic systems (so-called bulk metallic glasses and others), and network glass formers (including silicates, borates, phosphates, chalcogenides and halogenides). Their interparticle bond types vary from covalent, hydrogen, ionic, metallic to van-der-Waals bonds. Their glass-transition temperatures cover about one decade and their thermal-expansion coefficients vary by approximately 2.5 and 1.5 decades in the glass and liquid phases, respectively. In general, the available data basis is broader for the glass state than for the liquid phase.

The open symbols in Figs. 1d and e show the complete $\alpha(T_g)$ data sets for the liquid and glass states, respectively, using a double-logarithmic representation. The first conclusion from these figures is a clear correlation of the thermal expansion with the glass-transition temperature, namely a decrease of $\alpha_g$ and $\alpha_l$ with increasing $T_g$. Notably, this correlation holds across very diverse material classes (indicated by different symbols in the figures) with different bond types and drastically varying glass-transition temperatures. The scatter of the data certainly partly signals the fact that $\alpha$ was often measured employing very different techniques applied by various experimental groups during the last century. It probably also arises from variations in the width and separation from $T_g$ of the temperature regime where the thermal expansion was determined (see also Supplementary Notes 2 and 3).

As discussed above, in principle a decrease of $\alpha$ with increasing $T_g$, as demonstrated in Figs. 1d and e, is expected if a Lindemann-like scenario would apply for the glass-liquid transition, too. However, when assuming the validity of equations (2) and (3), such double-logarithmic plots of $\alpha$ versus $T_g$ (open symbols) should lead to approximately linear behaviour with slope -1. Instead, both data sets depend much stronger on $T_g$ as becomes obvious from a comparison with the upper solid lines, indicating slope -1, i.e. $\alpha \propto 1/T_g$. At best, only part of the liquid data, especially at $T_g < 400$ K, are roughly consistent with equation (3). This rationalizes the reported correlation of $\lambda$ with the fragility index $m$ (ref. [44]) within the framework of the KSZ model[41]. We find that an exponential $T_g$-dependent variation, $\alpha_i = \alpha_{0,i} \exp(-T_g / \theta_i)$ (with $i = g$ or $l$ for glass or liquid, respectively), as indicated by the dashed lines in Figs. 1d and e, provides a much better formal description of the experimental data than $\alpha_i \propto 1/T_g$ suggested by equations (2) and (3). Indeed, both data sets can be quite well linearized within a semi-logarithmic representation, plotting the logarithm of $\alpha_i$ versus $T_g$ (Supplementary Fig. 1). The only exception are the values for the borates in the liquid state, whose thermal expansion seems to represent a special case. Indeed, exceptional thermal-expansion properties of the borate glasses were identified earlier[19,48,49], and are believed to be due to their specific network structure.

The very similar $\alpha$-$T_g$ correlations for the liquid and glass state are astonishing, having in mind that the thermal expansion in the supercooled liquid includes vibrational as well as configurational contributions, while in the glass it should be dominated by vibrational contributions only.

Moreover, we find an approximately identical exponential factor $\theta_l = \theta_g \approx 270$ K, for both glasses and liquids. This implies a fixed ratio $\alpha_l/\alpha_g = \alpha_{0,l}/\alpha_{0,g}$. Using $\alpha_{0,l} \approx 1.4\times10^{-3}$ K$^{-1}$ and $\alpha_{0,g} \approx 4.3\times10^{-4}$ K$^{-1}$, obtained from the fits, this ratio is about three, which should be universally valid for all glass formers. To check this prediction, Fig. 2 shows $\alpha_l/\alpha_g$ versus $T_g$ for those materials where both expansion coefficients are available. Indeed, this ratio is close to three for a large variety of glass-formers belonging to different material classes. Only the borate glasses reveal much larger ratios, in accord with their known anomalous expansion behaviour[19,48,49].

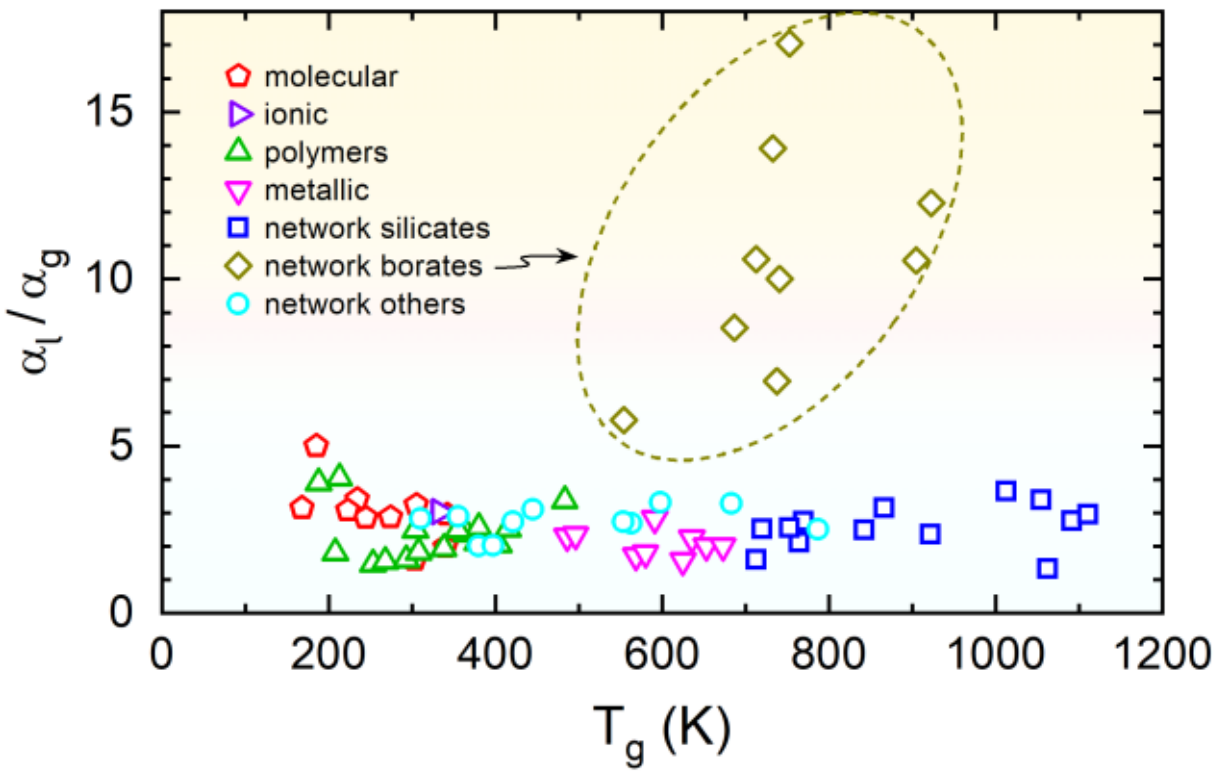

**Fig. 2 | Ratio of the thermal expansion coefficients measured in the liquid and glass phases.** Leaving the borate glasses aside, this ratio is of order three for all systems and independent of the glass-transition temperature.

## Discussion and Concluding remarks

We have shown that the thermal-expansion data of about 200 glass formers reveal a clear correlation with the glass-transition temperature, which holds across vastly different material classes. However, the data are clearly inconsistent with $\alpha T_g = \text{const.}$, expected when assuming a Lindemann-like scenario for the glass transition. This expectation is neither met for the glass, nor for the liquid phase, where it was theoretically predicted[40,41,44]. Instead, we find a much stronger decrease of $\alpha$ with $T_g$ for both states. This only becomes obvious when considering data covering a broad range of glass-transition temperatures and thermal-expansion coefficients.

The invalidity of equation (2) implies that at least one of the intuitive proportionalities $T_g \propto U_0$ and $1/\alpha_g \propto U_0$ (analogous to the crystal case; cf. introduction section and Supplementary Note 1) must be invalid for glasses. A clue is given when considering that $U_0$, the depth of the pair potential, essentially corresponds to the interparticle binding strength. As materials with very weak (van-der-Waals) and strong (covalent) bonds are included here, it should vary by about 2 - 3 decades. This is in accord with the observed variation of $\alpha_g$ (Fig. 1e), i.e. consistent with $1/\alpha_g \propto U_0$. In contrast, $T_g$ varies by 1.2 decades only and, thus, $T_g \propto U_0$ should be invalid. Therefore, we conclude that the transition temperature from glass to liquid depends much weaklier on the microscopic quantity $U_0$ than for the crystal-liquid transition where $T_m \propto U_0$. This marked difference seems to somehow reflect the fact that the glass transition qualitatively differs from crystal melting. This can be rationalized as follows:

Notably, the systems with small $T_g$ and high $\alpha$, lying in the upper left part of Fig. 1e (e.g., the polymers and molecular materials), generally exhibit higher fragility index $m$ than those with high $T_g$ and small $\alpha$ like the metallic or network systems[50] (cf. Supplementary Table 1). $m$ is a quantitative measure of the deviation of a material's viscosity $\eta$ from the Arrhenius temperature dependence, $\eta \propto \exp[E/(k_B T)]$, expected when assuming canonical thermally-activated particle dynamics with a well-defined energy barrier $E$ (refs. [42,43]). Such deviations are a hallmark feature of glass-forming liquids and strongly material dependent, being most pronounced, e.g., in many polymers and molecular liquids[1,4,43]. They are often ascribed to an increase of the effective energy barrier with decreasing temperature, caused by the cooperative motion of ever larger numbers of molecules upon cooling a liquid towards its glass transition[2,3,51]. Within this framework, higher $m$ values (characterizing so-called "fragile" glass formers[1,4,43]) mean that this increase is stronger than for small $m$ values ("strong" glass formers).

The stronger $T_g$ dependence of $\alpha_g$ compared to equation (2), observed in the present work, then could be due to this effective energy-barrier enhancement: The glass temperatures of the more fragile materials in the upper left part of Fig. 1e are larger than expected from their pair-potential depth alone, because, in order to liquify these glasses, more energy has to be invested to break up their cooperative particle network. Within this scenario, $T_g \propto m\, U_0$ instead of $T_g \propto U_0$ may be tentatively assumed. In contrast, the relation $\alpha_g \propto 1/U_0$ should be unaffected by cooperativity as thermal expansion is governed by the local pair potential only (Fig. 1b). Therefore, the proportionality $\alpha_g \propto 1/T_g$ should be invalid, in accord with experimental observation (Fig. 1e), and, instead, the quantity $\alpha_g/m$ should be proportional to $1/T_g$. This expectation indeed is well fulfilled as demonstrated by the filled symbols in Fig. 1e, showing $\alpha_g/m$ vs. $T_g$ for those systems where $m$ is known (cf. Supplementary Table 1). Notably, a corresponding cooperativity correction also is able to linearize the thermal expansion coefficients of the liquid state (filled symbols in Fig. 1d), i.e., we find $\alpha_l/m \propto 1/T_g$. Thus equations (2) and (3) should be replaced by:

$$\alpha_i/m\, T_g = const. \quad (i = \text{g,l}) \qquad (4)$$

Finally, it is remarkable that $\alpha_l$ and $\alpha_g$ (or $\alpha_l/m$ and $\alpha_g/m$) exhibit the same dependence on glass temperature and are related by a universal factor of about three, characterizing the increase of the thermal expansion when crossing the glass transition upon heating. A factor of 2 - 4 was occasionally quoted in literature[33,34] and here we document a factor close to 3 which is valid for the complete universe of glass-forming materials, leaving the borates aside. As discussed above, it is reasonable that the vibrational contributions to the thermal expansion are essentially the same in the glass and liquid states (cf. Fig. 1c), ascribing the observed higher $\alpha_l$ to additional configurational contributions arising above $T_g$ (refs. [33,34]). Then $\alpha_l/\alpha_g \approx 3$ implies that the configurational part is universally two times higher than the vibrational one, which seems surprising when considering their different physical origins. It is reasonable that the thermal expansion is related to the maximum possible displacement of a particle during the corresponding motion (either vibrational and/or configurational). If one expands the 3rd derivative of the pair potential versus distance (the thermal expansion coefficient) from one to three dimensions, assuming still the same local process, and adds the configurational – many body – motions, one can rationalize the detected factor of three. Thus, we conclude that the enhancement of $\alpha$ above $T_g$ is essentially a dimensionality effect. Locally we propose here the crossover from a two-body interaction (vibrations on the ps timescale) to an additional many-body process (configurational changes on a much longer time scale).

The found universal correlation of $\alpha_g$ and $T_g$, involving the degree of cooperativity of particle motion in different material classes, quantified by fragility *m*, obviously is a typical, so far unnoticed, property of glasses. It markedly differs from the much simpler behaviour of crystalline systems which can be explained in terms of the Lindemann criterion. This and the unexpected universal factor relating $\alpha$ in the glass to that in the liquid put severe constraints on existing and future models of the glass transition. Finally, the present results have predictive power for engineering glassy materials by design: one will be able to predict $T_g$ in a bottom-up way based on interatomic/intermolecular parameters and to deduce it from a simple thermal expansion measurement; conversely, a simple $T_g$ measurement will yield a wealth of information about atomic-scale composition and thermal properties.

## Data availability

The data that support the findings of this study are available in Supplementary Table 1.

## Acknowledgements

We thank Gyan Johari, Frank Stillinger and Dieter Vollhardt for stimulating discussions. K.S. acknowledges constant support over many years by the DFG Sa/337 via the Leibniz Program and Caltech, Pasadena, CA via the visiting associate program. B.R. is grateful for support from the VILLUM Foundation's Matter Grant (No. 16515). A.Z. gratefully acknowledges financial support from US Army Research Office, Contract No. W911NF-19-2-0055.

## Author Contributions

K.S. initiated this work. P.L. and B.R. collected the experimental data. P.L. analyzed the data and prepared the figures. A.L., P.L., K.S., and A.Z. wrote the manuscript. All authors discussed the results and commented on the manuscript.

## Competing interests

The authors declare no competing interests.

## Additional information

**Supplementary information** The online version contains supplementary material available at...

**Correspondence and requests for materials** should be addressed to Peter Lunkenheimer.

---

# Supplementary Information

for

# Thermal expansion and the glass transition

**Peter Lunkenheimer[1], Alois Loidl[1], Birte Riechers[2,3], Alessio Zaccone[4,5,6] and Konrad Samwer[7]**

[1] Experimental Physics V, Center for Electronic Correlations and Magnetism, University of Augsburg, 86135 Augsburg, Germany
[2] Bundesanstalt für Materialforschung und -prüfung, 12205 Berlin, Germany
[3] Glass and Time, Department of Science and Environment, Roskilde University, DK-4000 Roskilde, Denmark
[4] Department of Physics "A. Pontremoli", University of Milan, Via Celoria 16, 20133 Milan, Italy
[5] Department of Chemical Engineering and Biotechnology, University of Cambridge, Cambridge CB3 0AS, U.K.
[6] Cavendish Laboratory, University of Cambridge, Cambridge, CB3 0HE, U.K.
[7] 1. Physikalisches Institut, University of Goettingen, Germany

**Contents:**

### Supplementary Note 1: The Lindemann criterion and alternative melting rules for crystals

The first melting rule that gained considerable attention is generally assigned to Lindemann.[1] More than 100 years ago, he calculated the phonon frequencies of monoatomic materials from the melting temperature $T_m$ utilizing Einstein frequencies and assuming that at the melting temperature the atoms with a given vibrational amplitude touch each other. This Lindemann melting criterion was brought into his modern form mainly by work of Gilvarry[2] reformulating the melting criterion in terms of root mean-square (rms) phonon displacements and utilizing the Debye model in calculating the phonon eigenfrequencies. His main conclusion was that monoatomic crystals melt when their mean-square displacements reach a critical value of their vibrational amplitude which is of the order of 7 – 8 % of the next-nearest neighbor distance. Nowadays, the Lindemann criterion of melting usually implies a ratio of rms displacements of particles to the lattice spacing of order 0.05 – 0.16 (e.g., ref. 3).

The Lindemann melting criterion can be derived using phonon excitation in a strictly harmonic potential, while in any realistic melting process, anharmonic contributions to the pair potential, giving rise to thermal expansion, will play an essential role. In addition, thermal expansivity certainly can be much easier measured than rms displacements and early on there were attempts to derive melting rules including anharmonic lattice effects. The very first remarkable attempt was provided by Grüneisen[4]. Assuming a realistic lattice potential including attractive and repulsive interaction forces, he calculated a number of important thermodynamic quantities via a rather complete and probably the first equation of state for crystalline solids, where the dimensionless parameter $\gamma$ enters, which was later named Grüneisen-parameter. Amongst many other quantities, he calculated the volume expansion from 0 K to the melting temperature and derived the equation $(V_m - V_o)/V_0$ = const., where $V_m$ and $V_0$ denote the volume at the melting temperature and at 0 K, respectively. For monoatomic elements, Grüneisen found this constant characterizing the relative volume expansion from 0 K to the melting point to be of order ~ 0.08.[4] From this relation one can immediately derive the condition $\alpha_c T_m$ = const., with $T_m$ being the melting temperature and $\alpha_c$ the coefficient of thermal expansion in the crystalline state, assumed to be constant up to the melting temperature. The relation of Grüneisen's treatment and the Lindemann melting criterion were discussed in detail by Dugdale and MacDonald[5], by MacDonald and Roy,[6] and by Gilvarry.[2] These authors also documented that the derivation of Grüneisen is valid for different types of anharmonic potentials and that, in first respect, the Grüneisen-parameter describes the characteristic lattice anharmonicity. Remarkably, Grüneisen recalculated the mean amplitude of a single atom at the melting temperature and derived a ratio of 0.085 compared to the interatomic separation.

Taking the above given relation, $\alpha_c\, T_m$ = const., serious, implies that the involved quantities, the melting temperature as well as the coefficient of thermal expansion, both depend on the strength of the interatomic pair potential. One can expect that $T_m$ is proportional to the depth of the pair potential, $U_0$. Considering that $U_0$ essentially represents the binding energy and that interparticle bonds have to break up upon melting, provides a first simple argument in favor of this proportionality. Within the Lindemann melting scenario, a material becomes liquid when the vibration amplitude of the particles exceeds a certain limit. It is clear that for smaller $U_0$, i.e. a more shallow potential well, such a limiting vibration amplitude is reached already at lower temperature, compared to a potential with larger $U_0$ (cf. Fig. 1b of the main paper). Thus, within the Lindemann scenario it is plausible that $T_m \propto U_0$. On the other hand, it has been shown by MacDonald and Roy[6] that the thermal expansion is inversely proportional to the potential depth, $1/\alpha_c \propto U_0$, which can be explained by the fact that the attractive part of the pair potential becomes more shallow on decreasing $U_0$ as schematically visualized in Fig. 1b. This latter result was derived for different realistic interatomic potentials.[6] Hence, overall we then have $T_m \propto U_0$ and $1/\alpha_c \propto U_0$, leading to $\alpha_c T_m$ = const. Indeed, as outlined above, it has been documented by Grüneisen[4] that the Lindemann criterion (melting appears when the vibrational amplitude exceeds a critical ratio) and the relation $\alpha_c\, T_m$ = const., both follow from a generalized equation of state of the solid assuming a realistic anharmonic pair potential.

Experimentally, the latter relation has been checked for a number of material classes and the visualization of this relation has entered numerous textbooks and lecture notes. In literature, Straumanis[7] was the first to plot thermal expansivity vs. the melting temperature (however using degree Celsius) of cubic elements and noted the continuous decrease of thermal expansion vs. increasing melting temperatures. Shortly after, this curve was fitted by van der Reyden[8], however, again not using an absolute temperature scale. Finally it was documented by Bonfiglioli and Montalenti[9] that the melting points of metals scale with the inverse thermal expansion, as theoretically derived. As documented in the main text of the work, this relation later on was checked and analyzed in detail for various materials classes, e.g., by Van Uitert[10] and Granato *et al*.[11]

For completeness of this small introduction into the field of melting phenomena, it is important to notice the work of Born,[12] who proposed that a "rigidity catastrophe" caused by a vanishing elastic shear modulus determines the melting of crystalline solids. It has been shown by Jin *et al*.[13] that the melting of crystals probably is triggered by the Born and Lindemann criterion simultaneously.

## Supplementary Note 2: Extraction of thermal-expansion data from literature

The main emphasis of this work is the detailed analysis of the available and published experimental data basis of the temperature-dependent volume expansion coefficient $\alpha$ for a large variety of disordered materials, measured in the high-temperature supercooled liquid phase ($T > T_g$) as well as in the low-temperature glass state ($T < T_g$). The data basis on thermal expansivity and glass transition temperature, which has been analyzed in the course of this work and which is presented in Figs. 1d, e and 2 of the main text, is documented in Supplementary Table 1. A few facts are worth being mentioned here. For many disordered materials the published data basis is significantly broader than listed in Supplementary Table 1 and in many cases we had to select between different sources as will be discussed below (Supplementary Note 3). In first respect and whenever possible, we tried to provide reference to the original literature, avoiding numerous inaccuracies and mistakes that have been made during the past decades: Supplementary Table 1 always indicates the volume thermal expansion as defined in Eq. 3 of the main text; it is the temperature derivative of the volume normalized to the volume. In cases where the linear thermal expansion was measured in the original literature, we simply multiplied this number by a factor of 3, taking into account the isotropic character of liquids and glasses. As is clearly documented in Supplementary Table 1 and also can be seen in Fig. 1d and e of the main text, the published data basis is much broader for the thermal expansion in the glass state than in the supercooled-liquid phase, partly due to experimental problems of the high glass-transition temperatures in some silicate-derived network glass formers and also because often only solid-state dilatometry was used. The sometimes large scatter of published data, as documented in Fig. 1d and e, partly certainly is due to the application of different techniques with varying precision used to determine the thermal expansion. In addition, in determining the volume expansion coefficient below and above $T_g$, the temperature dependence of the volume has to be linearized and, of course, the deduced slope, used to calculate the thermal expansion, depends on the width of the temperature window investigated and on the distance or closeness to $T_g$. The temperature regime for linearization varies considerably in range and width in the different experiments reported. Whenever possible, we took the two thermal expansion coefficients and the glass transition temperature from the same reference. Certainly there exist many more experimental values of thermal expansion in literature. Specifically concerning the network glasses, many experiments reported on systematic investigations of glasses with various components as function of concentration. In these cases, we have chosen the most relevant data and in many cases used the compounds with the highest and lowest $T_g$ only, in order not to overburden the figures shown in the main part of the text.

## Supplementary Note 3: Choice of $\alpha$ and $T_g$ in case of multiple sources

In many cases, for a given material different values for the thermal expansion and the glass-transition temperature were reported and it is not straightforward to decide on the most reliable values. These variations result from different experimental techniques applied and/or differences in the samples investigated, e.g., different sample purity. Moreover, in polymers the chain length may vary considerably with considerable influence on thermal expansivity. In a number of molecular glasses, like alcohols or van-der-Waals glasses with low glass-transition temperatures, the water content may play a significant role and not always was specified in the published work. Concerning the glass-transition temperature, in addition different cooling rates and measuring and analyzing methods can also lead to considerable variations.

As an example, Supplementary Table 2 shows the various experimental reports on glass transition temperature and thermal expansivity for the molecular glass glycerol and for the chalcogenides network-glass selenium. Astonishingly, despite the fact that various techniques and various samples with mostly non-specified impurities were used, the scatter is within reasonable limits. An exception is the thermal-expansion coefficient of glycerol in the solid state, where the data from different sources deviate by more than a factor of three. In all cases, we decided not to take average values but to use the results with a clear description of the experimental measuring procedure, which are best documented, and from our point of view are most reliable.

## Supplementary Figure 1 (with discussion)

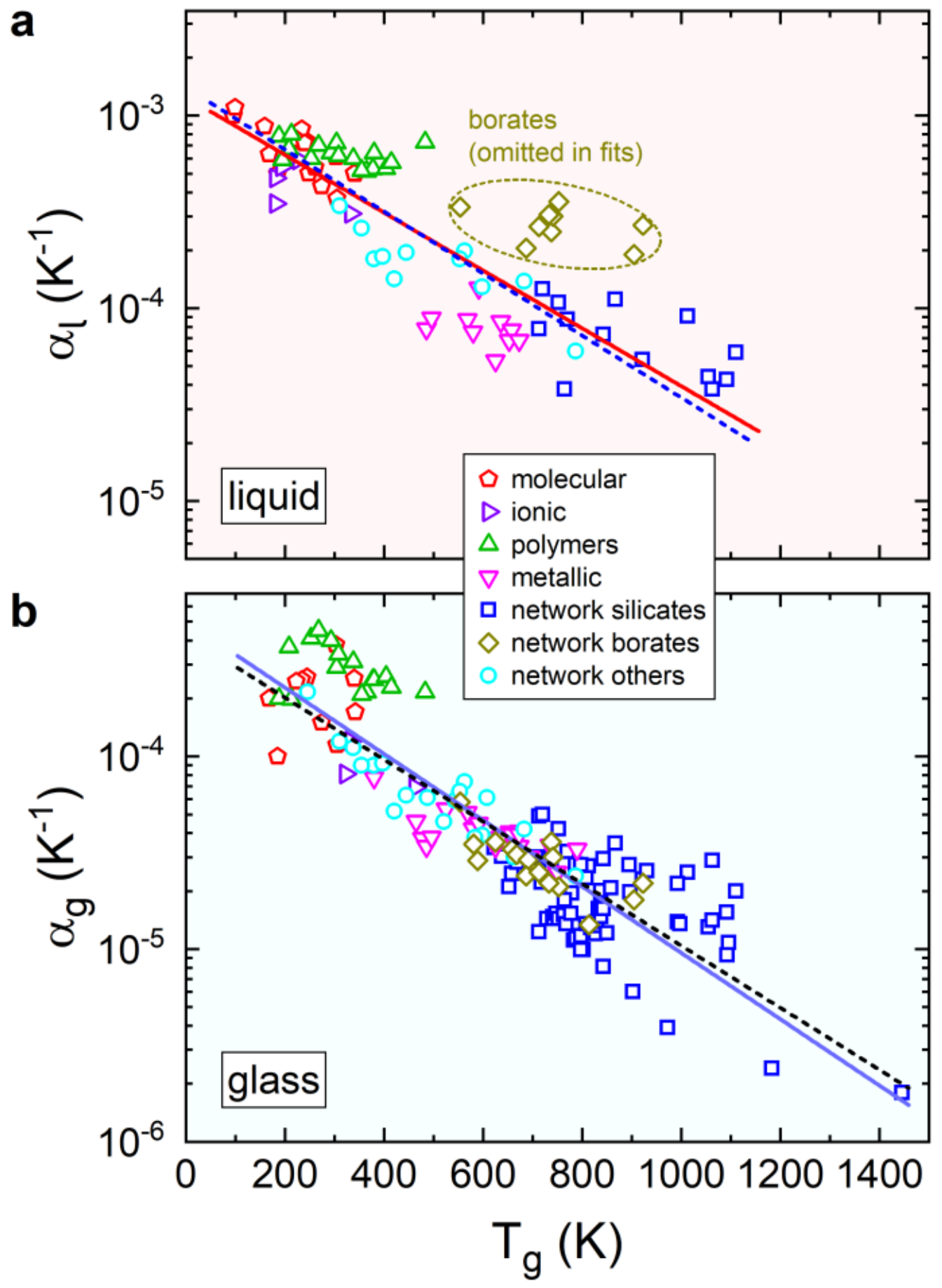

**Supplementary Figure 1 | Correlation of the thermal expansion with the glass-transition temperature.** Semilogarithmic plot of the thermal volume-expansion coefficients $\alpha_g$ in the glass phase (**a**) and of $\alpha_l$ in the liquid phase (**b**) versus the glass-transition temperature $T_g$ (same data sets as in Figs. 1d and e). The solid lines show linear fits based on all data points for each phase, except for $\alpha_l$ of the borates. The dashed lines represent linear fits presuming the same slope for $\alpha_l$ and $\alpha_g$. Note that the ordinates of (**a**) and (**b**) were adjusted to achieve the same decades/cm ratio, enabling a direct comparison of the slopes.

Supplementary Fig. 1 shows the complete thermal-expansion data sets for the liquid (a) and glass state (b) using a semi-logarithmic representation, $\log_{10} \alpha$ vs. $T_g$. The experimental values are the same as those shown in the double-logarithmic plots of Figs. 1d and e and as listed in Supplementary Table 1. The observed approximately linear decrease directly demonstrates the roughly exponential dependence of $\alpha$ on $T_g$ as also indicated by the dashed lines in Figs. 1d and e. The solid lines in Supplementary Fig. 1 represent linear fits, $\log_{10}\alpha_i = b_i - s_i T_g$ (with $i = g$ or $l$ for glass or liquid, respectively), neglecting $\alpha_l$ of the borates, which represent a special case (see main text). A simple visual comparison of frames (a) and (b) already reveals quite similar slopes $s$ for the liquid and glass data sets. Indeed, the free fits (solid lines) lead to slopes of comparable order, $s_g = 1.72\times10^{-3}$ K$^{-1}$ and $s_l = 1.50\times10^{-3}$ K$^{-1}$, and both data sets can be almost equally well fitted using the same slope of $1.61\times10^{-3}$ K$^{-1}$ as shown by the dashed lines (the latter fit curves are also shown by the dashed lines in Figs. 1d and e). Therefore, the $\alpha$-$T_g$ correlations of the liquid and glass phases only differ by their axis intercepts: $b_l = -2.85$ and $b_g = -3.37$. Overall, we then have $\alpha_i = \alpha_{0,i}\exp(-n\,T_g)$, with $\alpha_{0,i} = 10^{b_i}$ and the same $n = s$ ln10 for both data sets. Notably, if this exponential dependence would be an exact description of the experimental data, the values $\alpha_{0,l} \approx 1.4\times10^{-3}$ K$^{-1}$ and $\alpha_{0,g} \approx 4.3\times10^{-4}$ K$^{-1}$ would represent upper limits of the thermal expansion, which are approached for small $T_g$ values. The inverse of $n$ corresponds to a general energy scale of $\theta \approx 270$ K, or about 23 meV, describing the thermal expansion of both glasses and liquids, which universally obey $\alpha_i = \alpha_{0,i}\exp(-T_g/\theta)$. As mentioned in the main text, the finding of the same $\theta$ (or $n$) for the liquid and glass states implies a universal ratio $\alpha_l/\alpha_g \approx 3$ for all glass formers.

## Supplementary Table 1

**Supplementary Table 1 |** Glass temperature $T_g$, thermal volume expansion coefficient in the liquid ($\alpha_l$) and the glass state ($\alpha_g$), ratio $\alpha_l/\alpha_g$ , and fragility index $m$ for various materials belonging to different classes of glass formers. Only for part of the listed materials both expansion coefficients are available.

| GLASS FORMER | $T_g$ (K) | $10^4\alpha_l$ (K$^{-1}$) | $10^4\alpha_g$ (K$^{-1}$) | $\alpha_l/\alpha_g$ | $m$ |
|---|---|---|---|---|---|
| **molecular (alcohols)** | | | | | |
| sorbitol | 274 [14] | 4.3 [15] | 1.5 [15] | 2.9 | 93 [14] |
| xylitol | 248 [16] | 5.02 [17] | | | 86 [16] |
| glycerol | 185 [16] | 5.0 [18] | 1.0 [18] | 5.0 | 53 [14] |
| propylene glycol | 168 [16] | 6.3 [19] | 2.0 [19] | 3.2 | 52 [14] |
| 1-propanol | 96 [20] | 10 [21] | | | 40 [14] |
| ethanol | 99 [16] | 11 [22] | | | 52 [16] |
| **molecular (v.d. Waals)** | | | | | |
| ααβ-*tris*-naphthylbenzene | 342 [23] | 5.0 [23] | 1.7 [23] | 2.9 | 66 [14] |
| 1,1'-di(4-methoxy-5-methylphenyl)cyclohexane (BMMPC) | 261 [24] | 5.4 [25] | | | 72 [14] |
| ortho-terphenyl | 245 [26] | 7.34 [26] | 2.58 [26] | 2.8 | 81 [14] |
| 1,1'-bis(p-methoxyphenyl)cyclohexane (BMPC) | 240 [24] | 7.2 [25] | | | 96 [14] |
| 67mol% ortho-terphenyl - 33mol% ortho-phenylphenol | 234 [27] | 8.5 [27] | 2.5 [27] | 3.4 | |
| propylene carbonate | 159 [16] | 8.8 [28] | | | 104 [14] |
| **molecular (various)** | | | | | |
| sucrose | 340 [29] | 5.02 [29] | 2.54 [29] | 2.0 | 88 [30] |
| glucose | 305 [31] | 3.72 [31] | 1.15 [31] | 3.4 | 79 [32] |
| colophony | 303 [33] | 6.1 [33] | 3.8 [33] | 1.6 | |
| α-phenyl-o-cresol | 223 [34] | 7.53 [34] | 2.45 [34] | 3.1 | 83 [14] |
| salol | 218 [14] | 7.36 [35] | | | 73 [14] |
| **polymer** | | | | | |
| poly(2,6-dimethylphenylene oxide) | 483 [36] | 7.29 [36] | 2.16 [36] | 3.4 | 218 [37] |
| Polycarbonate | 415 [36] | 5.72 [36] | 2.28 [36] | 2.5 | 132 [14] |
| poly(ortho-methylstyrene) | 404 [38] | 5.31 [38] | 2.6 [38] | 2.0 | |
| poly(cyclohexyl methacrylate) | 380 [39] | 6.4 [39] | 2.5 [39] | 2.6 | |
| poly(methyl methacrylate) | 378 [40] | 5.3 [40] | 2.5 [40] | 2.1 | 145 [14] |
| polyvinyl chloride | 355 [41] | 5.2 [41] | 2.1 [41] | 2.5 | 191 [14] |
| Polystyrene | 365 [38] | 5.13 [38] | 2.16 [38] | 2.4 | 139 [14] |
| poly(ethyl methacrylate) | 338 [40] | 6.0 [40] | 3.1 [40] | 1.9 | 81 [42] |
| poly(n-propyl methacrylate) | 308 [40] | 6.2 [40] | 3.4 [40] | 1.8 | 63 [42] |
| polyvinyl acetate | 304 [43] | 7.2 [43] | 2.9 [43] | 2.5 | 95 [14] |
| poly(n-butyl methacrylate) | 293 [40] | 6.4 [40] | 4.0 [40] | 1.6 | 75 [44] |
| poly(n-hexyl methacrylate) | 268 [40] | 7.0 [40] | 4.5 [40] | 1.6 | |
| poly(n-octyl methacrylate) | 253 [40] | 6.0 [40] | 4.1 [40] | 1.5 | |
| Polyurethane | 213 [45] | 8.02 [45] | 1.98 [45] | 4.1 | |
| poly(n-dodecyl methacrylate) | 208 [40] | 6.7 [40] | 3.7 [40] | 1.8 | |
| polyisobutene | 195 [46] | 5.9 [46] | | | 46 [14] |
| polybutadiene | 188 [41] | 7.8 [41] | 2 [41] | 3.9 | 85 [32] |
| **ionic** | | | | | |
| $AgPO_3$ | 463 [47] | | 0.69 [48] | | |
| $[Ca(NO_3)_2]_{0.4}[KNO_3]_{0.6}$ (CKN) | 333 [49] | 3.64 [50] | 1.2 [51] | 3.0 | 93 [14] |
| $(AgI)_{0.67}(Ag_2MoO_4)_{0.33}$ | 323 [52] | | 0.81 [53] | | |
| 1-Butyl-3-methylimidazolium chloride (Bmim Cl) | 228 [54] | 5.9 [55] | | | 97 [54] |
| 1-Methyl-3-octylimidazolium hexafluorophosphate (Omim $PF_6$) | 194 [54] | 5.42 [56] | | | 78 [54] |
| 1-Butyl-3-methylimidazolium tetrafluoroborat (Bmim $BF_4$) | 182 [54] | 3.5 [57] | | | 93 [54] |
| 1-Butyl-3-methylimidazolium tetrachloroferrate (Bmim $FeCl_4$) | 182 [54] | 4.74 [58] | | | 144 [54] |
| **metallic** | | | | | |
| $Co_{43}Fe_{20}Ta_{5.5}B_{31.5}$ | 790 [59] | | 0.33 [59] | | |
| $Fe_{75}P_{16}Si_6Al_3$ | 750 [10] | | 0.25 [60] | | |
| $Fe_{65}Co_{10}Ga_5P_{12}C_4B_4$ | 735 [59] | | 0.345 [59] | | |
| $N_{75}P_{16}B_6Al_3$ | 695 [10] | | 0.30 [60] | | |
| $Co_{59}Ni_{10}Fe_5Si_{11}B_{15}$ | 673 [61] | 0.68 [61] | 0.34 [61] | 2.0 | |
| $Cu_{60}Hf_{25}Ti_{15}$ | 673 [59] | | 0.408 [59] | | |
| $Cu_{60}Zr_{30}Ti_{10}$ | 673 [59] | | 0.393 [59] | | |
| $Zr_{11}Cu_{47}Ti_{34}Ni_8$ | 658 [62] | 0.77 [63] | | | 47 [62] |
| $Pd_{76}Au_6Si_{18}$ | 658 [59] | | 0.393 [59] | | |
| $Ti_{41.5}Zr_{2.5}Cu_{42.5}Ni_{7.5}Hf_5Si$ | 654 [59] | | 0.405 [59] | | |
| $Zr_{55}Al_{10}Ni_5Cu_{30}$ | 653 [59] | | 0.339 [59] | | |
| $Zr_{65}Cu_{17.5}Al_{7.5}Ni_{10}$ | 653 [59] | 0.68 [63] | 0.339 [59] | 2.0 | |
| $Pd_{77.5}Cu_6Si_{16.5}$ | 636 [60] | 0.85 [60] | 0.38 [60] | 2.2 | 60 [64] |
| $Zr_{41.2}Ti_{13.8}Ni_{10}Cu_{12.5}Be_{22.5}$ | 625 [65] | 0.532 [65] | 0.339 [65] | 1.6 | 39 [64] |
| $Pd_{16}Ni_{64}P_{20}$ | 591 [66] | 1.27 [66] | 0.24 [66] | 2.8 | |
| $Pd_{48}Ni_{32}P_{20}$ | 580 [66] | 0.75 [66] | 0.42 [66] | 1.8 | |
| $Pd_{40}Ni_{40}P_{20}$ | 569 [67] | 0.87 [67] | 0.51 [67] | 1.7 | 50 [64] |
| $Pd_{42.5}Ni_{7.5}Cu_{30}P_{20}$ | 525 [59] | | 0.534 [59] | | |
| $Pt_{45}Ni_{30}P_{25}$ | 496 [66] | 0.89 [66] | 0.38 [66] | 2.3 | |
| $Pt_{52.5}Ni_{22.5}P_{25}$ | 486 [66] | 0.78 [66] | 0.34 [66] | 2.3 | |
| $Pt_{60}Ni_{15}P_{25}$ | 478 [59] | | 0.375 [59] | | |
| $La_{55}Al_{25}Ni_{20}$ | 465 [59] | | 0.459 [59] | | 37 [64] |
| $Mg_{65}Cu_{25}Y_{10}$ | 380 [59] | | 0.774 [59] | | 45 [62] |
| **network silicates** | | | | | |
| $SiO_2$ | 1446 [68] | | 0.018 [69] | | 20 [14] |
| 96.6$SiO_2$:2.9$B_2O_3$:0.4$Al_2O_3$:0.02$Na_2O$:0.02$K_2O$ wt% (Corning Vycor®) | 1183 [70] | | 0.024 [70] | | |
| 51.1$SiO_2$:25.2$Al_2O_3$:23.8CaO mol% (anorthite) | 1111 [71] | 0.59 [71] | 0.2 [71] | 3.0 | 54 [72] |
| 55.6$SIO_2$:22.2$A1_2O_3$:22.2MgO mol% (cordierite, melted) | 1096 [73] | | 0.108 [73] | | |
| 55.6$SIO_2$:22.2$A1_2O_3$:22.2MgO mol% (cordierite, sol-gel) | 1093 [73] | | 0.093 [73] | | |
| 53.4$SiO_2$:23.1$Al_2O_3$:19.1CaO:4.5$Na_2O$ mol% | 1092 [74] | 0.426 [74] | 0.155 [74] | 2.7 | |
| 60$SiO_2$:20$Al_2O_3$:20$Na_2O$ mol% | 1063 [75] | 0.38 [75] | 0.29 [75] | 1.4 | |
| Corning Jade® glass (alkaline earth aluminosilicate) | 1055 [76] | 0.44 [76] | 0.13 [76] | 3.4 | 32 [77] |
| 49.8$SiO_2$:25.6CaO:24.6MgO mol% (diopside) | 1013 [74] | 0.91 [74] | 0.25 [74] | 3.6 | 53 [72] |
| 69.0$SiO_2$:18.9$Al_2O_3$:12.3$Na_2O$ wt% (albite) | 922 [78] | 0.54 [78] | 0.23 [78] | 2.1 | 22 [72] |
| 65.8$SiO_2$:28.6$K_2O$:5.6$Al_2O$ mol% | 867 [74] | 1.11 [74] | 0.354 [74] | 3.1 | |
| 60$SiO_2$:26.6$Na_2O$:13.3$Al_2O_3$ mol% | 843 [75] | 0.73 [75] | 0.29 [75] | 1.4 | |
| 71.4$SiO_2$:14.3MgO:14.3$Na_2O$ mol% | 813 [79] | | 0.27 [79] | | |
| 82$SiO_2$:12$B_2O_3$:5$Na_2O$:1$Al_2O_3$ mol% (Duran) | 803 [80] | | 0.099 [80] | | |
| 68.4$SiO_2$:31.2PbO:0.4($Al_2O_3$ or $Fe_2O_3$) mol% | 779 [81] | | 0.2 [81] | | |
| 80$SiO_2$:20$Na_2O$ mol% | 770 [82] | 0.876 [82] | 0.32 [82] | 2.7 | 37 [14] |
| 90$SiO_2$:10$Na_2O$ mol% | 765 [82] | 0.38 [82] | 0.18 [82] | 2.1 | |
| 70$SiO_2$:30$Na_2O$ mol% | 753 [82] | 1.07 [82] | 0.42 [82] | 2.5 | 34 [72] |
| 59.8$SiO_2$:39.9PbO:0.3($Al_2O_3$ or $Fe_2O_3$) mol% | 733 [81] | | 0.23 [81] | | |
| 60$SiO_2$:40$Na_2O$ mol% | 720 [82] | 1.26 [82] | 0.50 [82] | 2.5 | 33 [72] |
| 66.7$SiO_2$:33.3$Na_2O$ mol% ($Na_2Si_2O_5$) | 713 [68] | 0.78 [83] | 0.49 [83] | 1.6 | 45 [14] |
| 71.4$SiO_2$:14.3CuO:14.3$Na_2O$ mol% | 694 [79] | | 0.27 [79] | | |
| 49.8$SiO_2$:49.8PbO:0.4($Al_2O_3$ or $Fe_2O_3$) mol% | 693 [81] | | 0.27 [81] | | |
| 39.4$SiO_2$:60.0PbO:0.6($Al_2O_3$ or $Fe_2O_3$) mol% | 623 [81] | | 0.34 [81] | | |
| **network silicates** (Schott technical glasses from ref. [84]) | | | | | |
| Schott 8240 (alkaline earth aluminosilicate glass) | 1063 [84] | | 0.141 [84] | | |
| Schott 8241 (alkaline earth aluminosilicate glass) | 1063 [84] | | 0.141 [84] | | |
| Schott NEO 1730 (alkaline earth, Nd containing, aluminosilicate glass) | 998 [84] | | 0.135 [84] | | |
| Schott 8252 (alkaline earth aluminosilicate glass) | 993 [84] | | 0.138 [84] | | |
| G018-346 (alkali-free sealing glass) | 993 [84] | | 0.219 [84] | | |
| Schott 8228 (sealing glass) | 973 [84] | | 0.039 [84] | | |
| G018-358 (alkali-free sealing glass) | 931 [84] | | 0.255 [84] | | |
| Schott 8229 (sealing glass) | 903 [84] | | 0.06 [84] | | |
| Schott 8436 (alkali alkaline earth silicate) | 897 [84] | | 0.198 [84] | | |
| Schott G018-311 (alkali-free sealing glass) | 895 [84] | | 0.273 [84] | | |
| Schott G018-266 (sealing glass) | 858 [84] | | 0.207 [84] | | |
| Schott 8341 (Pyran® S, borosilicate floatglass) | 850 [84] | | 0.121[84] | | |
| Schott 8450 (sealing glass) | 843 [84] | | 0.162 [84] | | |
| Schott 8230 (sealing glass) | 843 [84] | | 0.081 [84] | | |
| Schott 8455 (sealing glass) | 838 [84] | | 0.201 [84] | | |
| Schott 8326 (neutral glass) | 838 [84] | | 0.198 [84] | | |
| Schott 8412 (Fiolax® clear, borosilicate glass) | 838 [84] | | 0.147 [84] | | |
| Schott 8800 (neutral glass) | 838 [84] | | 0.165 [84] | | |
| Schott 8454 (sealing glass) | 838 [84] | | 0.192 [84] | | |
| Schott 8414 (Fiolax® amber, borosilicate glass) | 833 [84] | | 0.162 [84] | | |
| Schott G018-200 (Zn-B-Si passivation and solder glass) | 830 [84] | | 0.138 [84] | | |
| Schott G018-197 (Zn-B-Si passiviation glass) | 830 [84] | | 0.132 [84] | | |
| Schott 8660 (borosilicate glass) | 828 [84] | | 0.12 [84] | | |
| Schott 8651 (sealing glass) | 822 [84] | | 0.132 [84] | | |
| Schott 8488 (Suprax®, borosilicate glass) | 818 [84] | | 0.129 [84] | | |
| Schott 8449 (sealing glass) | 808 [84] | | 0.135 [84] | | |
| Schott 8415 (Illax®, soda-lime glass) | 808 [84] | | 0.234 [84] | | |
| Schott 8350 (AR-Glas®, soda-lime glass) | 798 [84] | | 0.273 [84] | | |
| Schott 8330 (Duran®, Borofloat® 33, Supremax®, borosilicate glass) | 798 [84] | | 0.099 [84] | | |
| Schott 8347 (borosilicate glass) | 798 [84] | | 0.099 [84] | | |
| Schott 8487 (borosilicate glass) | 798 [84] | | 0.117 [84] | | |
| Schott 8689 (borosilicate glass) | 788 [84] | | 0.114 [84] | | |
| Schott 8625 (VivoTag®, biocompatible glass) | 787 [84] | | 0.275 [84] | | |
| Schott 8448 (sealing glass) | 783 [84] | | 0.111 [84] | | |
| Schott 8245 (borosilicate glass) | 778 [84] | | 0.153 [84] | | |
| Schott 8652 (sealing glass) | 768 [84] | | 0.135 [84] | | |
| Schott 8250 (borosilicate glass) | 763 [84] | | 0.15 [84] | | |
| Schott 8270 (borosilicate glass) | 763 [84] | | 0.15 [84] | | |
| Schott 8447 (sealing glass) | 753 [84] | | 0.144 [84] | | |
| Schott 8650 (alkali-free sealing glass) | 748 [84] | | 0.153 [84] | | |
| Schott 8242 (borosilicate glass) | 743 [84] | | 0.144 [84] | | |
| Schott G017-725 (lead-borosilicate glass) | 741 [84] | | 0.147 [84] | | |
| Schott 8360 (soft glass, lead-free) | 738 [84] | | 0.273 [84] | | |
| Schott 8100 (lead glass, 33.5 % PbO) | 738 [84] | | 0.288 [84] | | |
| Schott 8405 (sodium-potassium-barium-silicate glass) | 733 [84] | | 0.291 [84] | | |
| Schott G017-096R (lead-borosilicate glass) | 729 [84] | | 0.144 [84] | | |
| Schott 8516 (sealing glass, high iron content) | 720 [84] | | 0.267 [84] | | |
| Schott 8456 (sealing glass) | 718 [84] | | 0.222 [84] | | |
| Schott 8337B (borosilicate glass) | 713 [84] | | 0.123 [84] | | |
| Schott 8470 (lead-free borosilicate glass) | 713 [84] | | 0.30 [84] | | |
| Schott 8531 (soft glass, sodium-free, high lead content) | 708 [84] | | 0.273 [84] | | |
| Schott 8532 (soft glass, sodium-free, high lead content) | 708 [84] | | 0.261 [84] | | |
| Schott 8095 (alkali-lead silicate, 28 % PbO) | 703 [84] | | 0.273 [84] | | |
| Schott G018-255 (lead-free Bi-Zn-B glass) | 669 [84] | | 0.282 [84] | | |
| Schott 8465 (lead-alumino-borosilicate glass) | 658 [84] | | 0.246 [84] | | |
| Schott G018-250 (lead-free solder glass) | 653 [84] | | 0.21 [84] | | |
| Schott G018-249 (lead-free solder glass) | 638 [84] | | 0.303 [84] | | |
| **network borates** | | | | | |
| 60$B_2O_3$:40CaO mol% | 923 [85] | 2.7 [85] | 0.22 [85] | 12 | |
| 75$B_2O_3$:25CaO mol% | 905 [85] | 1.9 [85] | 0.18 [85] | 11 | |
| Schott G018-205 (zinc-borate glass) | 814 [84] | | 0.134 [84] | | |
| 73$B_2O_3$:27PbO mol% | 753 [86] | 3.58 [86] | 0.21 [86] | 17 | |
| 60$B_2O_3$:40$Li_2O$ mol% | 741 [87] | 3.0 [87] | 0.30 [87] | 10 | |
| 66.7$B_2O_3$:33.3$Na_2O$ mol% | 738 [87] | 2.5 [87] | 0.36 [87] | 6.9 | |
| 66.7$B_2O_3$:33.3PbO mol% | 733 [86] | 3.06 [86] | 0.22 [86] | 14 | |
| 58$B_2O_3$:42PbO mol% | 713 [86] | 2.65 [86] | 0.25 [86] | 11 | |
| 50$B_2O_3$:25PbO:10$Na_2O$:15$Fe_2O$ mol% | 691 [88] | | 0.29 [88] | | |
| 85$B_2O_3$:15$Li_2O$ mol% | 687 [89] | 2.05 [89] | 0.24 [89] | 8.5 | |
| 50$B_2O_3$:25PbO:15$Na_2O$:10$Fe_2O$ mol% | 668 [88] | | 0.31 [88] | | |
| 50$B_2O_3$:25PbO:20$Na_2O$:5$Fe_2O$ mol% | 659 [88] | | 0.33 [88] | | |
| 50$B_2O_3$:25PbO:25$Na_2O$ mol% | 625 [88] | | 0.36 [88] | | |
| Schott G018-256 (lead-borate glass) | 589 [84] | | 0.288 [84] | | |
| Schott G017-052 (lead-borate glass) | 581 [84] | | 0.351 [84] | | |
| $B_2O_3$ | 554 [14] | 3.35 [90] | 0.58 [90] | 5.8 | 32 [14] |
| **network phosphates** | | | | | |
| 50$P_2O_5$:50BaO mol% | 683 [86] | 1.38 [86] | 0.42 [86] | 3.4 | |
| 45$P_2O_5$:45$K_2O$: 10$MgSO_4$ mol% | 608 [91] | | 0.61 [91] | | |
| 50$P_2O_5$:50$Na_2O$ mol% | 563 [86] | 1.98 [86] | 0.74 [86] | 2.7 | |
| 50$P_2O_5$:40$Na_20$:10BaO mol% | 553 [86] | 1.8 [86] | 0.66 [86] | 2.7 | |
| **network chalcogenides** | | | | | |
| $GeO_2$ | 787 [92] | 0.6 [92] | 0.24 [92] | 2.5 | 20 [14] |
| 50$B_2O_3$:33.3$TeO_2$:16.7PbO mol% | 678 [93] | | 0.31 [93] | | |
| 60Se:40Ge mol% | 598 [94] | 1.29 [94] | 0.39 [94] | 3.3 | |
| 60Se:35Ge:5Sb mol% | 584 [95] | | 0.38 [95] | | |
| 83.3$TeO_2$:10$LiO_2$:6.7PbO mol% | 543 [93] | | 0.59 [93] | | |
| 60Se:20Ge:20Sb mol% | 521 [95] | | 0.46 [95] | | |
| 70Se:20Ge:10Sb mol% | 488 [95] | | 0.61 [95] | | |
| 60Se:40As mol% | 445 [96] | 1.95 [96] | 0.63 [96] | 3.1 | 36 [42] |
| 60As:40Se mol% | 421 [96] | 1.42 [96] | 0.52 [96] | 2.7 | |
| 80Se:10Ge:10Sb mol% | 397 [94] | 1.86 [94] | 0.92 [94] | 2.0 [94] | |
| 80Se:20As mol% | 355 [96] | 2.6 [96] | 0.9 [96] | 2.9 | |
| Se | 310 [97] | 3.4 [96] | 1.2 [96] | 2.8 | 87 [14] |
| S | 246 [10] | | 2.16 [10] | | 86 [98] |
| **network halogenides** | | | | | |
| $BeF_2$ | 663 [10] | | 0.3 [10] | | 24 [42] |
| $ZnCl_2$ | 380 [10] | 1.8 [99] | 0.9 [10] | 2.0 | 30 [14] |
| 60$BeF_2$:40KF mol% | 338 [10] | | 1.11 [10] | | |

## Supplementary Table 2

**Supplementary Table 2** | Glass temperature $T_g$, thermal volume expansion coefficient in the liquid ($\alpha_l$) and the glass state ($\alpha_g$), and the ratio $\alpha_l/\alpha_g$ for glycerol and selenium taken from a variety of different sources as indicated in the table. The bold numbers are those given in Supplementary Table 1 and shown in the figures of the main text.

| GLASS FORMER | $T_g$ (K) | $10^4\alpha_l$ (K$^{-1}$) | $10^4\alpha_g$ (K$^{-1}$) | $\alpha_l/\alpha_g$ |
|---|---|---|---|---|
| **glycerol** | **185 [16]** | **5.0 [18]** | **1.0 [18]** | **5.0** |
| | ~ 184 [18] | | | |
| | | ~ 4.5 [100] | ~ 0.85 [100] | 5.3 |
| | | 5.0 [101] | 1.0 [101] | 5.0 |
| | 183 [102] | 5.0 [102] | 0.9 [102] | 5.6 |
| | 185 [103] | 5.4 [103] | 3.3 [103] | 1.6 |
| | 180 [104] | 4.8 [104] | 2.4 [104] | 2.0 |
| | 180 - 190 [29] | 4.83 [29] | 2.4 [29] | 2.0 |
| **Se** | **310 [97]** | **3.4 [96]** | **1.2 [96]** | **2.8** |
| | | | 1.4 [105] | |
| | 304 [102] | 4.0 [102] | 1.3 [102] | 3.1 |
| | 302 [103] | 4.6 [103] | 1.7 [103] | 2.7 |
| | 303 [33] | 4.46 [33] | 1.74 [33] | 2.6 |
| | 323 [94] | 3.54 [94] | 1.32 [94] | 2.7 |
| | 300 [104] | 4.2 [104] | 1.7 [104] | 2.5 |
| | 323 [96] | | | |
| | 308 [106] | | | |
| | 304 [107] | | | |